\documentstyle[preprint,tighten,aps]{revtex}

\def\lapprox{\mathrel{\mathop
  {\hbox{\lower0.5ex\hbox{$\sim$}\kern-0.8em\lower-0.7ex\hbox{$<$}}}}}
\def\gapprox{\mathrel{\mathop
  {\hbox{\lower0.5ex\hbox{$\sim$}\kern-0.8em\lower-0.7ex\hbox{$>$}}}}}
\def\permille{$^\circ/_{\circ \circ} $}

\def\diu{$u/U_{mod}$}
\def\gam{$\gamma$}
\def\yfo{y$_{ph}$}
\def\ug{$U$}
\def\N_F{$n_F$}
\def\nnu{\Omega}
\def\oomega{\omega}

\begin{document}

\preprint{\vbox{\noindent
          \null\hfill  INFNFE-20-96}}
\title{Helioseismology and standard solar models}
\author{
         S.~Degl'Innocenti$^{1,2}$,
         W. A. Dziembowski$^{3}$,
         G.~Fiorentini$^{1,2}$,
         and B.~Ricci$^{2}$
       }
\address{
$^{1}$Dipartimento di Fisica dell'Universit\`a di Ferrara, 
       via Paradiso 12, I-44100 Ferrara, Italy\\
$^{2}$Istituto Nazionale di Fisica Nucleare, Sezione di Ferrara, 
      via Paradiso 12, I-44100 Ferrara, Italy\\
$^{3}$Copernicus Astronomical Center, ul. Bartycka 18, 00716 Warsaw, Poland
        }

\date{November 1996}
\maketitle                 
\begin{abstract}
We present a systematical analysis of  uncertainties in the helioseismological
determination of quantities characterizing the solar structure. We discuss the effect
of errors on the measured frequencies, the residual solar
model dependence and the uncertainties of the inversion
method. We find Y$_{ph}=0.238-0.259$, $R_b/R_\odot=0.708-0.714$ and
$\rho_b=(0.185-0.199)$ gr/cm$^3$ (the index $b$ refers to the bottom of the
convective envelope). In the interval $0.2<R/R_\odot<0.65$, the quantity
\ug=$P/\rho$ is determined with and accuracy of $\pm 5$\permille~or better.
The predictions of a few recent solar model calculations are compared with helioseismological
results.
\end{abstract}

\section{Introduction}
\label{intro}

A Standard Solar Model (SSM) may be defined as a description
of the solar interior   which reproduces 
the observed properties of the sun within observational errors, by adopting
a set of physical and chemical inputs chosen within the range
of their uncertainties.

From the point of view of stellar evolution, the sun is a well known
structure. Confidence in   SSMs rests on the
success of stellar evolution theory to describe many, and more complex, 
evolutionary phases in good agreement with observational data.
Before the advent of helioseismology, however,  a solar model 
had three (essentially) free parameters -- the initial helium
and heavier elements abundances, Y and Z, and
the mixing length parameter $\alpha$ --  for producing three
observables: the present radius, luminosity and
heavy elements content of the photosphere.
From this point of view, the success of SSMs to reproduce 
these observables may  look as a not too big accomplishment.

The solar neutrino puzzle (see e.g. \cite{libro,Bere,Report}),
 i.e. the disagreement between the SSM 
predictions and the results of the four
 solar neutrino experiments\cite{Cloro,Kamioka,Sage,Gallex},
originated many discussions about the validity of the SSMs and prompted many
attempts to build alternative solar models, none however capable
of solving the discrepancy.

In recent years helioseismology has added important information on  the 
solar structure. A SSM  has  now to account for these
additional data.

Helioseismology (see \cite{Science} 
for an introduction and \cite{Oscill,ARAA}
for extensive reviews)
 provides severe constraints and 
tests of solar model calculations.
For instance, from inversion (deconvolution) of helioseismological data one can
 infer the photospheric helium abundance Y$_{ph}$ \cite{Vor91,DJ91}
 and the 
depth of 
the convective zone $R_b$ \cite{sismo}.
The helium mass fraction, which was 
used
as an  essentially  free parameter to match the observed solar luminosity, is now 
strongly constrained. Still one  can adjust  $\alpha$ in 
order  to  get the proper solar radius, $R_\odot$;  however there is no parameter which 
can be tuned to get $R_b$. The 
comparison  between the theoretically predicted value of $R_b$ and the value 
inferrred from helioseismology is a test of the model.

Helioseismology can tell a lot  about the solar interior, 
however, it cannot replace  SSMs. 
As an example,  helioseismology  determines  
the sound speed  down  to the solar core to high accuracy, but 
it cannot give the temperature profile in the energy production region.
In order to calculate the   several neutrino luminosities (pp, $^7$Be, 
CNO, $^8$B...), which crucially depend on this profile,   SSMs are
needed.  

For people interested in solar neutrinos, 
the basic question is thus:

\begin{itemize}
\item
Which solar models -- if any -- are acceptable, i.e. consistent with helioseismology?
\end{itemize}

Before addressing this question, one has to  provide a quantitative answer
to the following one:

\begin{itemize}
\item
How accurate are  the  helioseimological  determinations of quantities 
characterizing the solar interior?
\end{itemize}

The present paper is essentially devoted to this issue: we perform
a systematic and possibly exhaustive investigation of the uncertainties of the 
helioseismological approach, in order to estimate the global error to be assigned to 
helioseismological determinations of the solar properties.
 With this spirit, we  analyse  the helioseismological 
determinations of several physical quantities $Q$ characterizing the solar structure.
Concerning the outer part of the sun, we  discuss 
Y$_{ph}$,  $R_b$, the  sound speed $c_b$ and density  $\rho_b$ at the 
bottom of the convective zone.
Then  we  consider  the ``intermediate" solar interior ($x$=$R/R_\odot =0.2-
0.65$), analysing the behaviour of the squared isothermal sound speed,
\ug=$P/\rho$.
Finally we  investigate the inner region ($x \leq 0.2$), where 
nuclear energy  and neutrinos are produced.

For each quantity $Q$  we  determine the partial errors 
corresponding 
to each  uncertainties of the helioseismological method.
This approach -- which will be  better elucidated in sections \ref{inver} and 
\ref{error} -- is 
needed since helioseismology  measures {\em  only } the frequencies  
\{$\nnu$\} of solar p-modes, and  quantities characterizing the solar 
structure are indirectly inferred from the \{$\nnu$\}'s, through an inversion 
method.  Schematically, the procedure is the following:

a)One starts with a solar model, giving values $Q_{mod}$ 
 and predicting a set  \{$\nnu_{mod}$\} of frequencies.  
These  will be somehow different from the measured frequencies, 
$\nnu_\odot \pm \Delta \nnu_\odot$

b)One  then searches for  the corrections  $q$ to the solar model which 
are needed in order to match the corresponding frequencies
 \{$\nnu_{mod} + \oomega$\}  with the observed 
frequencies \{$\nnu_{\odot}$\}. 
Expression for $\oomega=\oomega(q)$ are derived 
 by using 
perturbation theory, where  the starting model is used as a zero-th order 
approximation. 
The corrections factor $q$ are then computed, assuming some regularity
 properties, so that
 the problem is  mathematically well defined and/or 
unphysical solutions are avoided.

c)The ``helioseismological value'' $Q_\odot$ is thus determined by adding 
the starting value and the correction:
\begin{equation}
Q_\odot= Q_{mod} + q  \, .
\end{equation}

There are three independent sources of uncertainties in this process:

i)Errors on the measured frequencies, which -- for a given inversion 
procedure -- propagate on  the value of $Q_\odot$.

ii)Residual dependence on the starting model:
the resulting $Q_\odot$ is slightly different
if one starts with  different   solar models. This introduces an 
additional uncertainty, which can be evaluated by 
comparing the results of several calculations.

iii)Uncertainty in the regularization procedure. Essentially this is a problem of 
extrapolation/parametrization. Different methods, equally acceptable in 
principle, yield (slightly) different values of $Q_\odot$.

It has to be remarked that, in view of the extreme precision of the measured 
frequencies, $ \Delta \nnu_{\odot} / \nnu_{\odot} \lapprox 10^{-4}$ 
\cite{ref1,ref2,ref3,ref4}, 
uncertainties corresponding to ii) and iii) are extremely important.
We consider all these uncertainties, as well as the possible ways of 
combining them to  estimate  the global error.

All this will provide a framework for a quantitative comparison between
solar models and helioseismology.  In the final part of the paper we test the 
predictions of a few recent (standard) solar model calculations. We feel that similar
test should be performed, when alternative solar models are proposed.

The paper is organized as follows:

A short review of the inversion methods is given in section \ref{inver},
 where we also define the free
parameters and determine their allowed ranges. 
The individual uncertainties mentioned above are operationally defined in 
section \ref{error}, where possible definitions of the global errors are also
discussed. In section \ref{outer} the properties of the outer sun (photospheric helium 
abundance 
and quantities characterizing the bottom of the convective zone) are examined, 
and  results are compared with previous analyses.
Section \ref{inter} is devoted to the intermediate region ($x =0.2-
0.65$) whereas section \ref{energy} concerns the energy production region.  
In section \ref{model} the results of a few recent  solar models are 
compared with  the available  helioseismological constraints/tests.

 The  concluding section describes the main results of the paper, summarized 
in Figs. \ref{fig1} and \ref{fig2} and in Table \ref{tab1}.

\section{Helioseismic data and inversion methods: a summary}
\label{inver}

 a) {\underline {Observational data}}\\

The helioseismic data that have been used for probing the solar interior 
structure are frequencies of acoustic oscillations (p-modes) corresponding to 
$Y_l^0$ spherical harmonics,  with the degrees $l$ ranging from 0 to 
some 150. At low degrees there are typically 15 modes corresponding to 
rather high radial degrees, $n>10$.
The total number of modes which has been used in structural 
inversions exceeds two thousands, which is only a small 
fraction of the total number of modes with measured 
frequencies. The latter number is higher by two orders of magnitude. However, 
modes  corresponding  to $l>150$ are not really useful for probing the
structure of the sun's interior, because they propagate only in the
shallow outermost layers and also because the accuracy of the
corresponding frequencies 
is much worse. Tesseral ($m\ne0$) mode frequencies yield an 
information about the sun's internal rotation and magnetic field. In 
the present 
application, the only relevance of these data is to justify ignoring
effects of the centrifugal and Lorentz forces on the radial structure.

The most exploited helioseismic data set is one from measurements at 
the Big Bear Solar Observatory in 1986 (BBSO86 \cite{ref1}).  This set contains no 
data for modes with $l<3$ and has been supplemented with data on
$l=0$--3 modes from various sources. The most extensive data on such modes
is from the network of automatic telescopes operated by the group of 
Birmingham University (BISON \cite{ref2}). Recently, data covering the whole range of 
$l$-values became available from the LOWL instrument \cite{ref3} and 
from the GONG network \cite{ref4}.\\

b){\underline{ Expression for $\oomega=\oomega(q)$}}\\

The basic assumptions made in seismic probings
of the sun's interior structure include 
hydrostatic equilibrium and  adiabaticity of oscillations.
In early inversions an asymptotic approximation for the radial 
eigenfunctions was used leading to a simple integral 
equation connecting the adiabatic sound speed, $c(R)$, to oscillation
frequencies $\nnu_{l,n}$. However, this 
approximation is inadequate, especially for probing the deep interior and in
most of the subsequent works numerical solutions have been adopted.
The price is a need to use a SSM as a starting model about which  
the hydrostatic equations are linearized. There is an implicit assumption that
the model provides a sufficently close description of the sun's structure.
With this assumption, the  variational principle following from the equation 
for the adiabatic oscillations, leads to an integral equation connecting the 
differences between the solar and the model 
functions describing 
radial structure (e.g. Ref. \cite{ref5}) to the corresponding differences in mode 
frequencies. 
The equation, however, cannot be directly applied to infer the difference 
in the structural functions because the assumption of hydrostatic 
equilibrium of the mean structure as well as that of adiabaticity of 
oscillations break down near the surface. To account for these departures 
an {\it ad hoc} term
of the form $F(\nnu)/ I$, where $\nnu$ is the frequency an
$I$ is
a suitable moment of inertia, must be added .
Such a form
follows from the fact that there the radial eigenfunctions are
nearly $l$-indepedent.
The resulting formula may be written in the following
form,
\begin{equation}
\label{X1}
\left ( \frac{\nnu_{\odot} - \nnu_{mod}}{\nnu_{mod}}  \right )_j
=\int{\cal K}_{Q,j}\frac{q}{Q_{mod}}dx+
\int{\cal K}_{\Gamma,j}\frac{{\mbox{\gam}}}{\Gamma_{mod}}dx+
\frac{F(\nnu_{mod \, j})}{I_j},
\end{equation}
where $j\equiv(l,n)$ identifies the mode, $Q(x)$ is a
structural function, $\Gamma$= ($d$logP/$d$log$\rho)_{ad}$
 is the adiabatic exponent, 
$x=R/R_{\odot}$ and:
\begin{eqnarray}
& q=&  Q_{\odot} -Q_{mod} \\ 
& {\mbox{\gam}}=&  \Gamma_{\odot} - \Gamma_{mod} \nonumber 
\end{eqnarray} 
There is freedom in choosing $Q$. It could be density $\rho$,
pressure $P$, or any
combination of these quantities and their derivatives.
All such functions
are connected through the linearized mechanical
equilibrium condition.

Even with a very large number of Eqs. (\ref{X1}), equal to the number 
of all modes in  the data set, a determination of the three
functions: $q(x)$, \gam($x$)~and
$F(\nnu)$ is not possible without the additional assumption that the 
unknown functions vary slowly.
One may eliminate one of the functions
by making use of a thermodynamical relation 
 $\Gamma=\Gamma(P,\rho,Y)$, where $Y$ is the fractional He abundance,
and of two, well justified simplifications, which are:\par
(1)  Y=Y$_{ph}$ in the convective envelope, and\par
(2)  \gam=0 in the radiative interior.\par 
With these additional constraints the 
unknown function \gam~is related with
the unknown number Y$_{ph}^{\odot}$.
This approach was first adopted in \cite{ref6} with the choice that the single
directly determined function was $U=P/\rho$.

Thus Eqs. (\ref{X1}) are written in terms of the unknown function
\begin{equation}
u(x)=U_{\odot} - U_{mod}, \nonumber
\end{equation}
the function  $F(\nnu)$ and of the unknown number
 \yfo~= Y$_{ph}^{\odot}-$Y$_{ph}^{mod}$.
One advantage of this formulation of 
the inverse problem is that one obtains directly the surface abundance of
He. We follow this approach here.  
Inversion without reference to thermodynamics has been done by 
Antia and Basu \cite{ref7}.\\

c) {\underline{ The regularized least square method and its free parameters}}\\

One way of making use of the assumption about the slow variability 
of \diu~and $F(\nnu)$ is a discretization in terms
of known functions. Following \cite{ref6}, we use cubic splines in the
 first case and Legendre polynomials in the second
case.
The coefficients in these two representations together
with \yfo~
are determined by the least-squares method. A
simple-minded application
of this method results in a solution for  \diu~
which exhibits artificial oscillations. A cure to
this problem is regularization (see e.g. \cite{ref8})
which consists in adding to the usual $\chi^2$ an
additional term.
Here we minimize
\begin{equation}
\label{X2}
\chi_{\rm reg}^2=\sum_{j=1}^J \left ({\nnu_{\odot} - \nnu_{mod}\over
\Delta \nnu_{\odot}} \right )_j ^2+{\lambda\over J}
\int \left ({d^2\over dx^2}{u\over
U_{mod}} \right ) ^2dx
\end{equation}
where J is the total number of the frequency data, 
$\Delta\nnu_{\odot}$ are the errors on the measured frequencies
and $\lambda$ is a control parameter. 
This form of regularization is referred to as the second derivative
smoothing. Alternative forms of regularization and influence of the choice of 
$\lambda$ were investigated in \cite{ref9}.

The regularization
parameter $\lambda$ was chosen to be the minimum value that still
suppresses the oscillations. Numerical experiments were conducted in which
attempt was made to reproduce known differences in
$u(x)$ between
two solar models from differences in p-mode frequencies.
The set of modes was the same as in the observational
data with weights
determined by the observational errors.
The conclusion from these experiments was that for $x>0.1$
it is possible to reproduce  $u(x)$ very accurately and
independently
of the choice of the type of the regularization. The
results are also
insensitive
to $\lambda$ in a wide range of values. 
However in the inner core ($x<0.1$) a regularization needed to
avoid the oscillatory
behaviour always flattened the real steep increase in \diu. 
It became clear that the method is not applicable
for probing this part of the sun. The reason is that
only a small fraction of modes in the data sets exibits any measurable
sensitivity to the sound speed changes in the inner core and the effect of
regularization dominates over the data.

Uncertainities on \yfo~and $u(x)$, as well as on other indirectly determined
structure functions, following from observational errors, are best estimated
by random number simulation having normal distribution with the 
half-width given by $\Delta\nnu_{\odot}$. There is an uncertainty  
following from the freedom in: choice of the regularization parameter
$\lambda$, number of bins, $n_b$, and number of the Legendre polynomials,
\N_F. This has been discussed in \cite{ref6,ref9,ref10}. Changes in $n_b$ may always
be compensated  by changes in $\lambda$ and will not be discussed 
here. Admissible ranges for $\lambda$ and \N_F~follow from considering the
behavior of $\chi^2_{reg}$.
 An increase in $\lambda$, as seen in Eq. (\ref{X2}), causes an increase 
in $\chi^2_{reg}$, the same effect being caused by the decrease of \N_F. Specifying
the maximum departure from the minimum value
$\chi^{2 ~(MIN)}_{reg} $
 we can set an upper bound for $\lambda$
and a lower bound for \N_F. 
Concerning $\lambda$, one adds the requirement of avoiding
 artificial oscillations. 
Dealing with the real data we may still distinguish
the artificial oscillations from features implied by data. The former have 
the half-wavelength equal to the bin length and their amplitude increases
sharply with the decrease of $\lambda$. Appearence of the oscilllation
sets a lower bound for $\lambda$.

 Let us summarize the role of
 the parameters of this method, which will be referred to as the 
{\it regularized least square} (RLS) method:\\
i) There are two parameters, $\lambda$ and \N_F.\\
ii) The seismic model of the sun is obtained by taking 
$\lambda$=0.001, as this is the minimal value for which unphysical
oscillations are avoided, and \N_F=20 as this corresponds to the minimal
degree of the polynomial such that $\chi^2_{reg}$ is stable.\\
iii)For this set of parameters, by using the ``model S'' of Ref. \cite{sunjcd}
(hearafter referred to as JCD), as a starting model, we obtain
 $\chi^{2~(MIN)}_{reg}$=2.5, which provides an indication of the fit
quality.\\
iv) For the parameters  $\lambda$ and \N_F~we consider as acceptable
those values which yield a   $\chi^2_{reg}$  not substantially
degraded with respect to the minimum. As a prescription borrowed
from statistics we accept the parameters if 
$\chi^2_{reg} \leq \chi^{2~(MIN)}_{reg} +1$. Numerically we found
that this implies $\lambda \leq 0.1$ and \N_F$\geq$10.\\

d) {\underline {Methods and parameters for the solar core}}\\
 
As already remarked, the method described above 
 cannot be used for probing the core within
$x$=0.1.
However, a fairly accurate seismic sounding of that region should 
be possible. 

It was found [23] that frequencies for some 30 solar p-modes 
were changed by more than $\Delta \nnu_{\odot}$ when $U$ in this innermost 
part of the core was changed by just one percent.
For that region {\it the optimal 
localized averaging} (OLA) method
is a better choice.
In this method we do not try to determine
functions. Instead, we try to determine mean values  weighted with
(possibly narrow) kernels centered at selected \{$x_0$\}
values. We consider a
linear superposition of individual kernels for the modes
present in the data set,
\begin{equation}
\label{X3}
K_u(x_0,x)=\sum_j^J c_j(x_0){\cal K}_{u,j}(x)
\end{equation}
and determine the coefficients $c_j$ at selected
points.
An application of the  classical method of Backus \&
Gilbert \cite{ref11}
to helioseismic data was described by Gough \& Thompson
\cite{ref5}.
The {\it subtractive optimal localized averaging}
(SOLA) method adopted in \cite{ref9} and used also in this paper was
developed by
Pijpers \& Thompson \cite{ref12}. Here one tries to construct
kernels $K_u(x_0,x)$ which are as close as possible to
Gaussians, $G$, centered in $x_o$ and
characterized by their half-widths at half-maximum, $w$.
Subject to minimization is the quantity
\begin{equation}
\label{X4}
\int \left [ K_u-G \left ({x-x_0\over w}\right ) \right ]^2dx+
\mu\sum_j^Jc_j ^2\Delta\nnu_{\odot j}^2.
\end{equation}
The second term, with a trade-off parameter $\mu$, is
added to avoid large error magnification.
For a specified $w$, larger $\mu$ leads to smaller errors
in the localized
mean but the kernels may differ significantly from
Gaussians
and therefore $w$ cannot be regarded as the measure of
localization.
Once the coefficients $c_j$ are determined the localized averages are:
\begin{equation}
\label{X5}
<{ u \over U } >=\sum_j c_j \left ( \nnu_{\odot}- \nnu_{mod}\over\nnu_{mod}  \right ) _{j}\pm
\sqrt{\sum_jc_j^2\Delta\nnu _{\odot j}  ^2 }
\end{equation}
A comparison of results of inversions with the two methods, 
given in \cite{ref9,ref10}, shows a very good agreement
everywhere except for the inner core.
Results obtained with the SOLA method are more realistic because 
it does not introduces an artificial smoothing.
Thus, for $x>0.1$ we may rely  
on the functional form of $u(x)$ as obtained by
means of the
RLS method. However, for a reliable
and accurate probing
of the inner core a different approach is required.
In a {\it hybrid} method of seismic model construction
developed in \cite{ref9}, the RLS method is used to
determine $F(\nnu)$, \yfo~and \diu~in the $[x_f,1]$ interval, whereas
the SOLA method is used to determine a single average value in the 
inner core around the point $x_0$.
 The effect of  $F(\nnu)$and \yfo~is removed from the data.
The function \diu~in the $[0,x_f]$ range
is then represented as a three term power series in
$x^2$ thereby satisfying the boundary condition at $x=0$. 
The three coefficients in the series are determined using the 
average SOLA value with the true kernel around $x_0$ and the continuity 
conditions at $x=x_f$ for \diu~and 
its derivative.
The method introduces its own uncertainties resulting from 
some freedom in choosing $x_0$ and  $x_f$.
 Regarding the role of the parameters of the hybrid method we
point out that:\\
i)The constraint on $x_f$ is that the point should be
located in the region where the SOLA values 
agree within the observational errors with the values inferred by the
RLS method. This occurs for $x_{fit} \geq 0.075$. We shall
explore the effects of varying $x_f$ in the range [0.075 -- 0.125].\\
ii)We shall also study the effect of varying $x_0$, while
$x_f$ is kept at the central value $x_f$=0.1. The range for
$x_0$ is suggested by two conditions: at too small
values there is not enough information. On the other hand,
when $x_0$ is too large and gets close to $x_f$ there is 
no arm for extrapolating the information to small
distances. As a consequence, somehow empirically,
we chose $x_0$ in the range [0.04 -- 0.06] and we adopt
 $x_0=0.05$ as the ``best'' value.\\
iii)The choice of the trade-off parameter, $\mu$,
for the hybrid method should be such as to minimize errors in \diu~
implied by the measurement errors; we chose $\mu$=0.001.
\medskip

\section{Partial uncertainties and global errors}
\label{error}

We  shall estimate how  quantities $Q$ characterizing the solar structure
are changed when some input of the helioseismological determination are
varied, one at a time. As a reference, we consider the values  $Q_\odot$ 
obtained with 
the seismic model described in the previous section:   frequencies
are kept at the central measured values, 
 the  JCD  model is used as the 
starting
model, and the inversion parameters  are kept at the ``best'' values
$\lambda$ = 0.001, $n_F$=20, $x_{fit}$=0.1 and $x_0$=0.05. 
(The results of the seismic model are presented in Table \ref{tab1},
second column, and Table \ref{tavolona} again second column.)\\

1){\underline{Observational errors}}.
We consider the effect of varying the frequencies from
the central measured values, within their observational errors.
We took random choices of frequencies within $\pm\Delta\nnu_{\odot}$
and evaluated the corresponding values
of $Q$. In this way we found maximal (minimal) acceptable values, 
$Q_{sup}$ ($Q_{inf}$). We define the allowed range as:
\begin{equation}
(\delta Q)_{exp}=Q_{sup} - Q_{inf}
%
%
\end{equation}

2){\underline{Starting model}}.
For the central values of the observed frequencies and the best values of
the inversion parameters, we study the differences resulting from
using different starting models. We remind that JCD includes diffusion
of helium and heavier elements, uses the Livermore equation of state and
the Livermore opacity tables. To assess the relevance of these inputs,
we performed inversion by using:\\
a) a model (SUN24l) with the same equation of state and opacities, but without
diffusion (modified version of ``model 0'' from Ref.\cite{ref10} ).\\
b) a model (SUN24) with the same opacities, the MHD equation of state and 
without diffusion (``model 0'' from Ref. \cite{ref10} ).\\
Exactly as before, for each quantity $Q$ we evaluated the allowed range, 
$( \delta Q )_{mod}$.\\

3){\underline{Inversion method}}.
As explained in the previous section, two different inversion methods are 
adopted, depending on the value of the radial coordinate.
 For $x \geq 0.1$, we use the RLS method
and consider the effect of varying $\lambda$ and $n_F$ separately.
The parameter $\lambda$ is varied in the range 0.001--0.1, 
see section \ref{inver}.
The resulting  variation (maximum  minus minimum in the explored  range)
 induced on the quantity $Q$ will be denoted as 
$( \delta Q )_{\lambda}$. The other parameter  $n_F$ is varied in the
range 10-20, see again section \ref{inver},
and  the resulting variation is denoted as $( \delta Q )_{n_F}$.

For the inner region we use the hybrid method:
the function $U(x)$ is determined 
by the requirement that an interpolating 
polynomial matches the results 
of the SOLA method around $x_0$ with 
that of the RLS method at  
$x_{f}$. As discussed in section \ref{inver}, we consider
$x_f$ in the range [0.075 -- 0.125] and  $x_0$ in the range
[0.04 -- 0.06].

Although we are confident that our analysis is rather exhaustive, there is no 
well defined  rule as to translate the differences  $( \delta Q  )_k$ into a global 
error   $ \Delta Q  $. Let us present two, somehow extreme, attitudes:

a)The statistical approach. If one believes that, for each parameter, all possible 
values lie in the interval $Q_k^{inf} - Q_k^{sup}$,  one can interpret
 $\pm 1/2 (\delta Q )_k$, as a partial error. Furthermore, one can assume that 
the partial 
uncertainties somehow  compensate each other, and use the combination 
rule for the   statistical fluctuations of independent variables, which  (for 
Poisson-Gauss distribution) yields:

\begin{equation}
\label{stat}
%
(\Delta Q)_{sta} =  \pm \frac{1}{2} \sqrt { \sum_k  (\delta Q)_k ^2 }
\end{equation}

b)The conservative approach.  May be that the parameter variation was not 
exhaustive, and what we found as extrema are not really so, but actually are 
quite acceptable values. In view of this, let us double  the interval we found 
and interpret $\pm( \delta Q  )_k$, as partial errors. Furthermore, let us be 
really conservative assuming that errors add up linearly. In conclusion, this gives:

\begin{equation}
\label{cons}
%
(\Delta Q_{con}) =  \pm  \sum_k  | (\delta Q)_k |  \, .
\end{equation}

In the following sections, we shall present both error estimates.  The first one
is reasonable when most of the error corresponds to  the statistical uncertainty of 
the measured frequencies. Actually in our case the other errors (from the 
residual model dependence and/or inversion method) are generally dominant. 
Clearly they are of a more subtle nature, a sort of systematic errors. For this 
reason, we feel more confident with the conservative error estimate, Eq. (\ref{cons}).

Before concluding this section, let us briefly consider the question of the error 
correlation, which is important when considering the helioseismological 
implications on  two quantities $Q^{(1)}$ and $Q^{(2)}$ (e.g. Y$_{ph}$ and $R_b$).
If the respective errors are $\Delta Q^{(1)}$ and  $\Delta Q^{(2)}$ , we shall 
consider 
a solar model as acceptable if the predictions lie within the rectangle  
$Q^{(1)} \pm \Delta Q^{(1)}$, $Q^{(2)} \pm \Delta Q^{(2)}$,  i.e we shall 
treat  errors as 
uncorrelated, 
whereas actually there is a correlation: as an example, the extremum   
$Q^{(1)} + \Delta Q^{(1)}$ most likely corresponds to a set of parameters 
which does 
not allow -- say --  $Q^{(2)} - \Delta Q^{(2)}$.  The helioseismologically 
allowed area 
should be an ellipsis contained  {\em inside the rectangle }. By 
choosing the rectangle 
as the allowed area, one is thus overestimating the error, which again  corresponds 
to a conservative attitude.

\section{ Properties of the convective envelope}
\label{outer}

\subsection{The photospheric Helium abundance}

Knowledge of the helium abundance in the sun is of fundamental 
importance both to cosmology and to solar structure theory.
Unfortunately, it is impossible to measure it, even in the atmosphere,
by means of standard spectroscopy. A helioseismological method of
determining Y$_{ph}$ was first suggested by Gough\cite{Gough84} and developed 
in Refs. \cite{DapGough84} and \cite{DGT88}. The value of Y$_{ph}$ inferred
for the sun, however, was not published, because it was feared that it might
be too sensitive to systematic errors in the data that were available
at the time. In \cite{DJ91} an alternative method was developed. Unlike
the former one, it does not rely on the validity of p-mode asymptotics.
Instead, it requires the validity of linearization about a reference model.

Several helioseismological determinations of Y$_{ph}$ have been published
 since 1991.
The various estimates, which are collected in Table \ref{elios},
 often differ among each other by more than the quoted errors,
see Fig. \ref{fig0}.
Actually, these latter often reflect  only the uncertainties of the measured 
frequencies, whereas the extracted value  of Y$_{ph}$ depends on the 
inversion method and 
on the physical inputs, see the discussion in Refs. \cite{RCVD96}
 and \cite{Kosov92}. Most of the estimates in Table \ref{elios} rely on the
MHD equation of state.

 By omitting the highest  value by Dappen et al. \cite{Dappen91} (the same 
authors later published significantly
smaller values \cite{Kosov92}) the total range of helioseismological determinations is:
\begin{equation}
\label{yph}
{\mbox{Y}}_{ph} =0.226 - 0.260 \,.
\end{equation}
Our reference estimate is Y$_{ph}^{\odot}$=0.249 (we remark the we rely on the
Livermore equation of state, which should be more accurate than MHD).
Concerning the allowed range,
the result of our conservative  analysis, shown in Table \ref{tab1}, implies:
\begin{equation}
\label{yph2}
{\mbox{Y}}_{ph} =0.238 - 0.259 \,,
\end{equation}
which substantially overlaps with the previous range.
From the same table one sees that experimental errors are of minor relevance,
the value of Y$_{ph}$ being manly sensitive to  the choice of the
starting model; as discussed in Refs. \cite{RCVD96} and \cite{Kosov92},
 the effect of the equation of
state is dominant.

\subsection{The bottom of the convective zone}

As emphasized in Ref. \cite{sismo}, the transition of the temperature gradient between
being subadiabatic and adiabatic at the base of the solar convective zone
gives rise to a clear signature in the sound speed. Helioseismic measurements
of the sound speed therefore permit a determination of the location
of the base of the convective zone.

In Ref. \cite{sismo} the following ranges were derived:
\begin{equation}
\label{rb}
R_b/R_\odot=0.710-0.716
\end{equation}
\begin{equation}
\label{cb}
c_b=(0.221-0.225) \,{\mbox{Mm/s}},
\end{equation}
where the quoted intervals include the 
uncertainty resulting 
from the inversion technique.
The estimate of $R_b$ has been  confirmed in Refs. \cite{Cox93,RCVD96}.

Actually, within the present uncertainty $R_b$ and $c_b$ are not
independent \cite{sismo}. As well known, the lower part of the
convective zone is very close to being adiabatically stratified,
the adiabatic exponent $\Gamma$ being close to 5/3, hence
\begin{equation}
P \propto \rho^{5/3}.
\end{equation}

Furthermore, throughout the same region, in the hydrostatic equation one
can approximate $M_x$  (the mass within $x$) with the total mass:
\begin{equation}
\frac{dP}{dx}= - \,\frac{GM_{\odot}}{ R_\odot \, x^2}\rho
\end{equation}

In terms of $U=P/\rho$ the two equations give:
\begin{equation}
\frac{dU}{dx} = - \, \frac{2}{5}\frac{GM_{\odot}}{R_\odot \, x^2}\rho
\end{equation}
and consequently
\begin{equation}
\label{cbr}
U(x) \simeq \frac{2}{5} \frac{G M_\odot}{R_\odot}  ( x -1 )
\end{equation}
 which shows that $U$ (hence c=$\sqrt{5/3 \, U}$) is uniquely known
as a function of x. In view of the approximations which have been used
[$M_{x} \simeq$ M$_{\odot}$,  $\Gamma \simeq 5/3$], Eq. (\ref{cbr})
is accurate to the percent level, and this is adequate in comparison
with the error of the helioseismological determination of c$_b$, see
Eq. (\ref{cb}).

On the other hand, the density $\rho_b$ at the bottom of the convective
zone is an independent quantity. Actually, from the above
equations $\rho(x)$ in the convective zone is determined up
to a scaling factor. The helioseismological determination of
$\rho_b$ fixes such a factor. Thus $\rho_b$ is also an interesting quantity,
which however received little attention in the past.

We derived $\rho_b$ from $U(x)$ by integrating the hydrostatic
equilibrium equation:
\begin{equation}
\label{eqidro}
\frac{1}{\rho} \frac{ d\rho}{dx}=
- \frac{1}{U} \left [ \frac{ dU }{ dx} + \frac{ G M_x }{ R_\odot \, x^2} \right ] \, ,
\end{equation}
together with
\begin{equation}
\label{masscons}
\frac{dM_x}{dx} =4 \pi \rho R_\odot x^2 \, .
\end{equation}
Thus $\rho_b$ is sensitive to all the uncertainties of the inversion
method, including those of small $x$ region.

Our analysis, see again Table \ref{tab1}, yields:
\begin{equation}
R_b/R_\odot=0.708 - 0.714
\end{equation}
essentially confirming the result of \cite{sismo}.
Note that again the main sensitivity is to the starting  model.
Concerning the density, we find:

\begin{equation}
\rho_b=(0.185 - 0.199) {\mbox{ g/cm$^3$}}
\end{equation}
Also in this case, a  large fraction of the error comes from the starting model.

\section{The intermediate region}
\label{inter}

The essential output of helioseismology is the reconstruction 
of the sound speed profile. Our discussion is in terms of the
related quantity $U$; as well known, c$^2$=$\Gamma U$ 
and below the convective zone 
$\Gamma$ = 5/3 with an accuracy of 10$^{-3}$ or better. 

By using the RLS method, it is possible to derive  directly the profile 
of  $U$  as a function of the 
radial coordinate throughout all the sun, except for  the inner region
 ($x<0.1$).
The results (values of $U$, partial uncertaintes 
 and global errors), are shown in Table \ref{tavolona} and 
Fig. \ref{fig1} and are summarized in Table \ref{tab1}.

It is convenient to consider an intermediate solar region:
$0.2<x<0.65$. The upper limit is established by 
requiring that it is well below the transition to the convective zone, which we discussed
above. The lower limit is chosen so as to exclude the region of energy production 
(see next section).
For this  region the  following 
comments are relevant:

a)Each of the individual uncertainties nowhere exceeds 
2~\permille.  

b)Uncertainties from the accuracy on the measured frequencies 
are of minor relevance with respect to the residual model dependence and to
the sensitivity to the inversion parameters.

c)All in all, even with the most conservative estimates, the helioseismological
determination is extremely accurate:
$|\Delta U/U | \leq  5{\mbox{\permille}}$
throughout the explored region.

\section{The energy production region}
\label{energy}

As well known, most of the energy and of solar neutrinos originate 
from the innermost part  of the sun. According to SSM calculations,
see e.g. Refs. \cite{Report,BP95}, about  
94\% of the solar luminosity  and 93\% of the pp neutrinos are produced 
within $x<0.2$, the region which we analyse in this section.

Our results, summarized in Fig. \ref{fig1} and  Table \ref{tab1},
deserve the following comments:\\
i) In the region    0.1$<x<$0.2, where the RLS method
holds, the error 
intrinsic to the inversion method is dominant, and 
the global uncertainty, which  worsens as one is going deeper into the sun, 
reaches $\pm$ 1\%.\\
ii)At even smaller radii, where the hybrid method is used to
determine the function $U(x)$, a large
fraction of the error comes  from the choice of $x_{f}$ and of $x_0$,
 whereas the residual dependence on  $\lambda$ and $n_F$ is rather weak. 
The result is also essentially unsensitive to the starting model.
 On the other hand, errors on frequencies become important.
 This obviously corresponds to  the fact that p-modes do not penetrate 
in the solar core , and consequently the information one can
 extract from available experimental results is limited.\\ 
iii)We remind that the production of $^7$Be neutrinos 
 is limited to a small region close to the solar center,
 the maximal production
occurring at:
\begin{equation}
\label{Be}
x_{Be}= 0.06 
\end{equation}
according to \cite{Report}.
The $^8$B neutrinos are generated even closer to the center,
 the maximal production being at \cite{Report}:
\begin{equation}
\label{B}
x_{B} = 0.04.
\end{equation}
(The exact locations of these maxima are somehow model dependent:
 for instance, in the  solar  model with helium and metal
diffusion of Ref.  \cite{BP95}, hearafter referred to as BP95,
one has $x_{Be}$ $\simeq$ 0.055  and 
$x_{B}$ $\simeq$ 0.045).
 One sees from Table \ref{tab1} that $U(x_{Be}$) and $U(x_{B}$),
as derived from the hybrid method, are 
globally
 known with an accuracy of about 2\%. The average values around
$x_{Be}$ and $(x_{B})$ given by the SOLA method are in agreement,
within the quoted errors, with the results of the hybrid method.

\section{Helioseismology and standard solar models}
\label{model}

A few recent solar models will now be compared  with the helioseismological 
information derived in the previous sections.

In this respect, the predicted vs. observed properties of the convective envelope are 
particularly interesting, see Table \ref{tab2}  and Fig. \ref{fig2}. All solar 
models neglecting diffusion cannot account for  Y$_{ph}$ and for the properties 
of the bottom of the convective zone ($R_b$ and $\rho_b$). Models with 
diffusion of helium and heavier elements are much closer to the 
helioseismological determinations. Among these, BP95  and JCD  
yield values of Y$_{ph}$, R$_b$ and $\rho_b$ in good agreement with helioseismological
predictions.

For understanding what is going on, let us concentrate on our own solar calculations.
FRANEC96 \cite{Ciacio} is our most recent solar model, including diffusion of
He and heavier elements and using the Livermore EOS and the new Livermore opacities
for 19 elements \cite{newopa} according to the composition of \cite{GN93}.
If diffusion is switched off, the resulting model (FRANEC96-ND) is
grossly inconsistent with helioseismology. FRANEC96 looks much better in this
respect. Hovewer it slightly 
underestimates the depth of the convective zone; as this is too shallow, the 
density at the bottom is  too small  (and  correspondingly the sound speed is somehow 
underestimated). If we use the older Livermore opacities calculated for 12 elements,
the resulting model (FRANEC96-OLD) is in good agreement with helioseismology.

In Fig. \ref{fig3} we show the difference between $U$ as predicted by selected solar 
models and the helioseismological determination, normalized to this latter.
We remark that BP95 and JCD model are in agreement with helioseismology everywhere.

SUN24, the model without diffusion of Ref. \cite{ref10} yields a good profile
of $U(x)$. We remind, hovewer, that it fails in predicting the properties of 
the convective envelope. This shows that the two approaches (profile of $U$ and 
properties of the convective envelope) are complementary and both important.

FRANEC96  underestimates $U$ by about 1\% 
near the transition between the convective and radiative region, as already 
discussed. On the other hand, FRANEC96-OLD, as well as other models using
older versions of the Livermore opacities (BP95, JCD and SUN24) 
look better. This does not imply some deficiency of the latest opacities.
The important point, however, is that helioseismology can discriminate
between models based on opacities which differ just by a few percent. 
This is best seen in Fig.\ref{fig5} where the profile of $U$ according to FRANEC96 
and FRANEC96-OLD are compared, together with the relative differences $\Delta k/k$
between opacities, these latter being calculated at any point $x$ for the same
value of $\rho$, T and chemical composition (given by FRANEC96-OLD).

 In conclusion, it seems to us that all these models with diffusion
essentially pass the helioseimological tests. Similar conclusions have been
recently reached  in Ref. \cite{BPBCD}, where the inconsistency of mixed
solar models with helioseismology is also discussed.

\section{Concluding remarks}
\label{conc}

We summarize here  the main points of this paper:

\begin{itemize}

\item
An extensive  an possibly exhaustive investigation of
uncertainties in  helioseismic determinations
of solar properties has been performed.

\item
The accuracy on quantities characterizing the convective envelope
(Y$_{ph}$, $R_b$ and  $\rho_b$) has been studied, see Table \ref{tab1},
 as well as that on the squared isothermal sound  speed,  
$U=P/\rho$, along the solar profile, see Fig. \ref{fig1}.

\item
Only recent standard solar models including diffusion of He and heavier
elements predict properties of the convective envelope in
agreement with helioseismology, see Table  \ref{tab2} and  Fig. \ref{fig2}.

\item
The same models also predict  for $U(x)$ values consistent
with the helioseismic determination all over the solar profile, see Fig. \ref{fig3}.

\end{itemize}
Solar models built {\em ad hoc } so as to solve the solar neutrino
problem should be at least as successful.

\acknowledgments
 W.A.D. is grateful to the director and the staff of the 
 Osservatorio Astronomico di Collurania (Teramo) for their hospitality and
 he gratefully acknowledges the financial support
from Laboratori Nazionali del Gran Sasso of INFN.
We appreciate useful discussion with V. Berezinsky and V. Castellani.
We are grateful to F. Ciacio for his help in developing the version
of FRANEC including diffusion of helium and heavier elements.
We thank S. Basu, J. Christensen-Dalsgaard, A. Dar, R. Sienkiewicz 
and  S. Turck-Chi\`{e}ze for providing us with
detailed results of their solar models. 
We thank A.A. Pamyatnykh and P.R. Goode who collaborated with W.A.D. on 
development of the inversion codes used in present work.
This work was partially supported by Ministero dell' Universit\`{a} e della
Ricerca Scientifica.

\newpage

\newpage 

\begin{table}
\caption[cc]{Helioseismological determinations of the photospheric
He abundance, Y$_{ph}$. }
\begin{tabular}{ll}
  Ref.         &   ~~~~Y$_{ph}$  \\
\hline
Dappen (1991) \cite{Dappen91}  &    0.268 $\pm$ 0.010\\
Vorontsov (1991) \cite{Vor91}   &  0.250 $\pm$ 0.010 \\
Dziembowski (1991) \cite{DJ91}  &  0.234 $\pm$ 0.005 \\
Kosovichev (1992) \cite{Kosov92}&  0.232 $\pm$ 0.006 \\
Cox (1993) \cite{Cox93}       &    0.240 $\pm$ 0.005 \\
Dziembowski (1994) \cite{DJ}  &   0.24295 $\pm$ 0.0005\\
Hernandez (1994) \cite{hern}  &   0.242 $\pm$ 0.003  \\
Antia (1994) \cite{Antia94}   &   0.252 $\pm$ 0.003 \\
Gough (1996) \cite{Gough96}   &   0.248 $\pm$ 0.005\\
 RVCD96 \cite{RCVD96}         &   0.250 $\pm$ 0.005     
\label{elios}
\end{tabular}
\end{table}

\begin{table}
\caption[a]{
For the indicated quantities $Q$ we present the helioseismological 
reference values
$Q_\odot$, the relative partial uncertainties $(\delta Q/Q)_{k}$ corresponding 
to experimental errors ($exp$), model dependence ($mod$), and to the 
parameters used in the inversion method ($\lambda$,  $n_{F}$, $x_{fit}$  and 
$x_{o}$). Global errors $\Delta Q/Q$, estimated according to the
statistical ($sta$) and the conservative ($con$) approach, see Eqs. 
(\ref{stat}) and (\ref{cons}),
are also shown.
 All uncertainties and errors are in  \permille. 
In the  fifth and sixth row, 
for the quantity $U=P/\rho$ the values of the partial uncertainties 
and of the global errors are the maxima in the 
indicated interval.
In the last two rows the results on $U$ at points representative of the 
$^7$Be  and $^8$B
 neutrino production, Eqs. (\ref{Be})
and (\ref{B}), have been derived by using the hybrid 
 method.
} 
\begin{tabular}{c | c | c | c | c c  c  c| c | c  }
$Q$ &$Q_\odot$&$\left ( \frac{\delta Q}{Q}  \right ) _{exp} $ &
$\left ( \frac{\delta Q}{Q}  \right ) _{mod} $ &
$\left ( \frac{\delta Q}{Q}  \right ) _{\lambda} $ &
$\left ( \frac{\delta Q}{Q}  \right ) _{n_F} $ &
$\left ( \frac{\delta Q}{Q}  \right ) _{x_{fit}} $ &
$\left ( \frac{\delta Q}{Q}  \right ) _{x_{o}} $ &
$\left ( \frac{\Delta Q}{Q}  \right ) _{sta} $ &
$\left ( \frac{\Delta Q}{Q}  \right ) _{con} $ \\
\hline
%
Y$_{ph}$ &0.249& 2.4 & 27&6   & 7 & & &14 &  42 \\ 
%
$R_b/R_\odot$ &0.711& 0.1 & 4 & 0.03 & 0.01& & &2 & 4  \\ 
%
 $\rho_b$ [g/cm$^3$] & 0.192 & 3 &16 & 4 & 6.4 &2.5 & 4.8 & 9.4 & 37   \\ 
\hline
%
$U$($0.2<x<0.65$ )& & $<$ 1  &  2 &  1.3  & 1.4 & &  &1.4  &5 \\
$U$($0.1<x<0.2$ )&  & $<$1  &  1.6 &  2.5 & 3.0 & 1.5 & & 2.3  & 9.4  \\
\hline
$U$$(x_{Be})$ [$10^{15}$ cm$^2$ s$^{-2}$]& 1.56  & 2.7  & 1.5  & 1.7 & 2.1 &3.5 & 6.6 & 4.3  & 18  \\
$U$$(x_{B})$  [$10^{15}$ cm$^2$ s$^{-2}$] & 1.55 & 4.8  &  1.3 &  $<$1 & 1.5 & 3.7&11 & 6.6  &  24   \\
\end{tabular}
\label{tab1}
\end{table}

\begin{table}
\caption[cc]{ Helioseismological determinations and solar model predictions
for the properties of the convective envelope}
 \begin{tabular}{lcccc}
  model & Ref.  &$R_b/R_\odot$   &  Y$_{ph}$   & $\rho_b$ [g/cm$^3$]  \\
  \hline
 \hline
 \multicolumn{3}{l}{\bf Models without diffusion:}\\
    FRANEC96-ND &\cite{Ciacio}    & 0.728  & 0.261 & 0.156 $^{\,}$ \\
    BP95-ND     &\cite{BP95}      & 0.726  & 0.268 & 0.157 $^{\,}$\\
		  CGK89-ND				&	\cite{COX}      & 0.714  & 0.291 & \\
    P94-ND      &\cite{P94}       & 0.726  & 0.270 & \\
    RVCD96-ND   &\cite{RCVD96}    & 0.725  & 0.278 & 0.166 $^{\,}$\\
    DS96-ND     &\cite{DS}        & 0.731  & 0.285 & \\
    BCDSTT-ND   &\cite{Basu}      & 0.721  & 0.276 & 0.171 $^*$\\
    TCL         &\cite{TCL}       & 0.725  & 0.271 & 0.166 $^*$\\
    SUN24       &\cite{ref10}     & 0.716 $^*$  & 0.283 $^*$ & 0.184 $^*$\\
    SUN24l      &                 & 0.714  $^*$ & 0.282 $^*$ & 0.186 $^*$\\
  \hline
 \hline
  \multicolumn{3}{l}{\bf Models with He diffusion:}\\ 
  BP92 &   \cite{BP92}   & 0.707  & 0.247  & 0.197  $^{\,}$\\
  P94  &   \cite{P94}    & 0.710  & 0.246  &\\
  BCDSTT & \cite{Basu}   & 0.707  & 0.246  & 0.199 $^*$\\
  \hline
  \hline
 \multicolumn{3}{l}{ \bf Models with He and Z diffusion:}\\
 FRANEC96 &\cite{Ciacio} & 0.716  & 0.238    & 0.181  $^{\,}$\\
 FRANEC96-OLD &          & 0.713  & 0.242    & 0.187  $^{\,}$\\
 BP95     &\cite{BP95}   & 0.712  & 0.247    & 0.187  $^{\,}$\\
 CGK89    &\cite{COX}    & 0.721  & 0.256    &  \\
 P94      &\cite{P94}    & 0.712  & 0.251    &  \\
 RVCD96   &\cite{RCVD96} & 0.716  & 0.258    & 0.188  $^{\,}$\\
 DS96     &\cite{DS}     & 0.713  & 0.231    & \\
 JCD      &\cite{sunjcd} & 0.711  & 0.245    & 0.190 $^*$\\
\hline
\hline
{\bf Helioseismology} & & 0.708--0.714 & 0.238--0.259 & 0.185--0.199  \\
 \end{tabular}
{\footnotesize $^*$ private comunication by the authors}
\label{tab2}
\end{table}

\newpage
\begin{table}
\caption[xx]{
 We present as a function of  $x=R/R_{\odot}$ the helioseismological 
(best) values for $U=P/\rho$ 
($U_\odot$), the relative partial uncertainties $(\delta U/U)_{k}$ corresponding 
to experimental errors ($exp$), model dependence ($mod$), and to the 
parameters used in the inversion method ($\lambda$,  $n_{F}$, $x_{fit}$ and 
$x_{o}$). Global errors $\Delta U/U$, estimated according to the
statistical and the conservative approach, see Eqs. (\ref{stat}) and (\ref{cons}),
are also shown.}
\begin{tabular}{c|c|c|c|cccc|c|c}
$x$ &$U_\odot$ [c.g.s.] &$\left ( \frac{\delta U}{U}  \right ) _{exp} $ &
$\left ( \frac{\delta U}{U}  \right ) _{mod} $ &
$\left ( \frac{\delta U}{U}  \right ) _{\lambda} $ &
$\left ( \frac{\delta U}{U}  \right ) _{n_F} $ &
$\left ( \frac{\delta U}{U}  \right ) _{x_{fit}} $ &
$\left ( \frac{\delta U}{U}  \right ) _{x_{o}} $ &
$\left ( \frac{\Delta U}{U}  \right ) _{sta} $& 
$\left ( \frac{\Delta U}{U}  \right ) _{con} $ \\
\hline
0.0084&1.539E+15&6.829E-03&1.170E-03&1.589E-04&1.004E-03&1.315E-02&1.591E-02&1.090E-02&3.823E-02\\
0.0099&1.540E+15&6.792E-03&1.169E-03&1.617E-04&1.013E-03&1.296E-02&1.583E-02&1.080E-02&3.792E-02\\
0.0116&1.540E+15&6.740E-03&1.169E-03&1.733E-04&1.027E-03&1.269E-02&1.571E-02&1.067E-02&3.751E-02\\
0.0136&1.542E+15&6.670E-03&1.233E-03&1.901E-04&1.046E-03&1.233E-02&1.555E-02&1.050E-02&3.702E-02\\
0.0160&1.543E+15&6.571E-03&1.242E-03&2.126E-04&1.071E-03&1.183E-02&1.533E-02&1.026E-02&3.626E-02\\
0.0188&1.544E+15&6.438E-03&1.424E-03&2.434E-04&1.106E-03&1.116E-02&1.503E-02&9.940E-03&3.540E-02\\
0.0221&1.546E+15&6.257E-03&1.545E-03&2.855E-04&1.154E-03&1.025E-02&1.462E-02&9.511E-03&3.411E-02\\
0.0260&1.549E+15&6.010E-03&1.638E-03&3.426E-04&1.219E-03&9.045E-03&1.407E-02&8.945E-03&3.232E-02\\
0.0305&1.551E+15&5.681E-03&1.590E-03&4.825E-04&1.305E-03&7.473E-03&1.332E-02&8.216E-03&2.985E-02\\
0.0357&1.554E+15&5.249E-03&1.449E-03&6.643E-04&1.419E-03&5.490E-03&1.234E-02&7.324E-03&2.661E-02\\
0.0416&1.557E+15&4.698E-03&1.348E-03&8.941E-04&1.566E-03&3.107E-03&1.109E-02&6.322E-03&2.271E-02\\
0.0482&1.560E+15&4.026E-03&1.346E-03&1.172E-03&1.746E-03&5.675E-04&9.564E-03&5.344E-03&1.842E-02\\
0.0555&1.562E+15&3.248E-03&1.408E-03&1.488E-03&1.957E-03&2.164E-03&7.783E-03&4.578E-03&1.805E-02\\
0.0634&1.563E+15&2.407E-03&1.472E-03&1.823E-03&2.189E-03&4.245E-03&5.833E-03&4.127E-03&1.797E-02\\
0.0717&1.561E+15&1.579E-03&1.473E-03&2.144E-03&2.429E-03&5.052E-03&3.853E-03&3.726E-03&1.653E-02\\
0.0805&1.557E+15&9.057E-04&1.542E-03&2.406E-03&2.661E-03&4.142E-03&2.035E-03&3.056E-03&1.369E-02\\
0.0896&1.549E+15&6.468E-04&1.549E-03&2.550E-03&2.852E-03&2.977E-03&6.362E-04&2.584E-03&1.121E-02\\
0.0992&1.538E+15&7.289E-04&1.496E-03&2.495E-03&2.962E-03&1.809E-03&3.901E-06&2.293E-03&9.494E-03\\
0.1093&1.523E+15&8.151E-04&1.576E-03&2.207E-03&2.993E-03&7.635E-04&0.000E+00&2.095E-03&8.354E-03\\
0.1200&1.501E+15&9.119E-04&1.553E-03&1.755E-03&2.977E-03&8.726E-05&0.000E+00&1.949E-03&7.284E-03\\
0.1315&1.473E+15&9.455E-04&1.441E-03&1.178E-03&2.886E-03&0.000E+00&0.000E+00&1.781E-03&6.452E-03\\
0.1440&1.438E+15&8.597E-04&1.313E-03&5.731E-04&2.690E-03&0.000E+00&0.000E+00&1.583E-03&5.435E-03\\
0.1575&1.394E+15&7.009E-04&1.130E-03&3.038E-04&2.352E-03&0.000E+00&0.000E+00&1.360E-03&4.487E-03\\
0.1722&1.342E+15&6.796E-04&1.230E-03&3.305E-04&1.866E-03&0.000E+00&0.000E+00&1.179E-03&4.105E-03\\
0.1880&1.283E+15&7.318E-04&1.313E-03&1.559E-04&1.280E-03&0.000E+00&0.000E+00&9.900E-04&3.480E-03\\
0.2045&1.221E+15&6.849E-04&1.143E-03&3.633E-04&7.345E-04&0.000E+00&0.000E+00&7.821E-04&2.926E-03\\
0.2216&1.157E+15&6.705E-04&1.109E-03&8.623E-04&4.040E-04&0.000E+00&0.000E+00&8.040E-04&3.045E-03\\
0.2389&1.096E+15&6.653E-04&1.336E-03&1.106E-03&3.668E-04&0.000E+00&0.000E+00&9.469E-04&3.474E-03\\
0.2564&1.037E+15&6.130E-04&1.415E-03&1.118E-03&4.892E-04&0.000E+00&0.000E+00&9.832E-04&3.635E-03\\
0.2744&9.814E+14&5.931E-04&1.373E-03&1.025E-03&6.370E-04&0.000E+00&0.000E+00&9.612E-04&3.629E-03\\
0.2927&9.286E+14&5.502E-04&1.606E-03&8.315E-04&7.288E-04&0.000E+00&0.000E+00&1.013E-03&3.716E-03\\
0.3107&8.811E+14&5.701E-04&1.590E-03&3.112E-04&7.928E-04&0.000E+00&0.000E+00&9.459E-04&3.264E-03\\
0.3286&8.373E+14&5.031E-04&1.402E-03&1.665E-04&8.328E-04&0.000E+00&0.000E+00&8.573E-04&2.904E-03\\
0.3466&7.958E+14&4.220E-04&1.380E-03&2.092E-05&8.116E-04&0.000E+00&0.000E+00&8.277E-04&2.634E-03\\
0.3646&7.564E+14&4.168E-04&1.511E-03&8.028E-04&7.492E-04&0.000E+00&0.000E+00&9.567E-04&3.479E-03\\
0.3829&7.198E+14&4.011E-04&1.463E-03&1.111E-03&7.803E-04&0.000E+00&0.000E+00&1.018E-03&3.755E-03\\
0.4013&6.858E+14&3.739E-04&1.456E-03&7.498E-04&9.174E-04&0.000E+00&0.000E+00&9.570E-04&3.497E-03\\
0.4198&6.535E+14&3.830E-04&1.606E-03&7.479E-04&1.007E-03&0.000E+00&0.000E+00&1.037E-03&3.744E-03\\
0.4385&6.229E+14&3.573E-04&1.878E-03&8.761E-04&9.760E-04&0.000E+00&0.000E+00&1.159E-03&4.087E-03\\
0.4575&5.942E+14&3.248E-04&2.010E-03&5.610E-04&9.858E-04&0.000E+00&0.000E+00&1.165E-03&3.882E-03\\
0.4768&5.670E+14&3.628E-04&2.076E-03&3.094E-04&1.107E-03&0.000E+00&0.000E+00&1.200E-03&3.856E-03\\
0.4963&5.404E+14&2.796E-04&1.966E-03&1.122E-03&1.106E-03&0.000E+00&0.000E+00&1.268E-03&4.474E-03\\
0.5160&5.154E+14&2.913E-04&1.775E-03&1.169E-03&1.070E-03&0.000E+00&0.000E+00&1.199E-03&4.305E-03\\
0.5360&4.916E+14&2.744E-04&1.787E-03&8.637E-04&1.140E-03&0.000E+00&0.000E+00&1.153E-03&4.065E-03\\
0.5562&4.686E+14&2.798E-04&1.987E-03&6.411E-04&1.199E-03&0.000E+00&0.000E+00&1.212E-03&4.106E-03\\
0.5765&4.464E+14&3.116E-04&1.916E-03&4.761E-04&1.179E-03&0.000E+00&0.000E+00&1.160E-03&3.882E-03\\
0.5968&4.248E+14&2.691E-04&1.953E-03&9.881E-04&1.298E-03&0.000E+00&0.000E+00&1.279E-03&4.508E-03\\
0.6172&4.040E+14&2.478E-04&1.544E-03&1.260E-03&1.238E-03&0.000E+00&0.000E+00&1.180E-03&4.290E-03\\
0.6374&3.837E+14&2.150E-04&1.871E-03&1.497E-03&1.288E-03&0.000E+00&0.000E+00&1.365E-03&4.872E-03\\
0.6574&3.638E+14&2.108E-04&1.861E-03&8.875E-04&1.423E-03&0.000E+00&0.000E+00&1.257E-03&4.383E-03\\
0.6769&3.435E+14&2.547E-04&1.593E-03&1.351E-04&1.479E-03&0.000E+00&0.000E+00&1.096E-03&3.462E-03\\
0.6957&3.218E+14&2.014E-04&2.392E-03&1.294E-03&1.622E-03&0.000E+00&0.000E+00&1.586E-03&5.508E-03\\
0.7131&2.991E+14&1.749E-04&2.170E-03&1.990E-03&1.821E-03&0.000E+00&0.000E+00&1.733E-03&6.155E-03\\
0.7298&2.751E+14&2.010E-04&2.095E-03&1.692E-03&2.014E-03&0.000E+00&0.000E+00&1.684E-03&6.002E-03\\
0.7461&2.529E+14&1.657E-04&2.068E-03&1.216E-03&2.155E-03&0.000E+00&0.000E+00&1.614E-03&5.604E-03\\
0.7619&2.321E+14&1.512E-04&2.196E-03&1.448E-03&2.272E-03&0.000E+00&0.000E+00&1.740E-03&6.067E-03\\
0.7772&2.127E+14&1.886E-04&2.334E-03&1.872E-03&2.344E-03&0.000E+00&0.000E+00&1.903E-03&6.738E-03\\
0.7920&1.946E+14&1.557E-04&2.381E-03&1.845E-03&2.390E-03&0.000E+00&0.000E+00&1.924E-03&6.771E-03\\
0.8063&1.779E+14&1.642E-04&2.641E-03&1.843E-03&2.575E-03&0.000E+00&0.000E+00&2.063E-03&7.224E-03\\
0.8199&1.623E+14&1.664E-04&3.086E-03&2.035E-03&2.850E-03&0.000E+00&0.000E+00&2.335E-03&8.137E-03\\
0.8329&1.478E+14&1.779E-04&3.271E-03&2.179E-03&2.969E-03&0.000E+00&0.000E+00&2.464E-03&8.597E-03\\
0.8453&1.345E+14&1.844E-04&3.385E-03&2.268E-03&3.035E-03&0.000E+00&0.000E+00&2.542E-03&8.872E-03\\
0.8571&1.222E+14&1.939E-04&3.700E-03&2.426E-03&3.281E-03&0.000E+00&0.000E+00&2.756E-03&9.600E-03\\
0.8682&1.109E+14&1.957E-04&4.087E-03&2.643E-03&3.659E-03&0.000E+00&0.000E+00&3.046E-03&1.058E-02\\
0.8786&1.005E+14&2.100E-04&4.209E-03&2.839E-03&3.945E-03&0.000E+00&0.000E+00&3.217E-03&1.120E-02\\
0.8884&9.099E+13&2.245E-04&4.166E-03&3.020E-03&4.157E-03&0.000E+00&0.000E+00&3.309E-03&1.157E-02\\
0.8975&8.229E+13&2.475E-04&3.969E-03&3.261E-03&4.460E-03&0.000E+00&0.000E+00&3.404E-03&1.194E-02\\
0.9060&7.435E+13&2.748E-04&3.865E-03&3.611E-03&4.958E-03&0.000E+00&0.000E+00&3.628E-03&1.271E-02\\
0.9139&6.712E+13&3.047E-04&3.810E-03&4.032E-03&5.591E-03&0.000E+00&0.000E+00&3.941E-03&1.374E-02\\
0.9211&6.054E+13&3.342E-04&3.663E-03&4.463E-03&6.215E-03&0.000E+00&0.000E+00&4.245E-03&1.467E-02\\
0.9278&5.457E+13&3.595E-04&3.464E-03&4.866E-03&6.722E-03&0.000E+00&0.000E+00&4.500E-03&1.541E-02\\
0.9340&4.915E+13&3.843E-04&3.496E-03&5.239E-03&7.195E-03&0.000E+00&0.000E+00&4.785E-03&1.631E-02\\
0.9397&4.424E+13&4.141E-04&3.814E-03&5.604E-03&7.701E-03&0.000E+00&0.000E+00&5.134E-03&1.753E-02\\
0.9449&3.980E+13&4.487E-04&4.083E-03&5.977E-03&8.294E-03&0.000E+00&0.000E+00&5.509E-03&1.880E-02\\
0.9496&3.578E+13&4.883E-04&4.579E-03&6.364E-03&8.998E-03&0.000E+00&0.000E+00&5.972E-03&2.043E-02\\
0.9540&3.215E+13&5.357E-04&5.507E-03&6.788E-03&9.830E-03&0.000E+00&0.000E+00&6.583E-03&2.266E-02\\
0.9579&2.887E+13&5.969E-04&6.528E-03&7.357E-03&1.085E-02&0.000E+00&0.000E+00&7.328E-03&2.533E-02\\
0.9616&2.591E+13&6.713E-04&7.725E-03&8.151E-03&1.212E-02&0.000E+00&0.000E+00&8.269E-03&2.867E-02\\
0.9649&2.323E+13&7.554E-04&9.347E-03&9.197E-03&1.365E-02&0.000E+00&0.000E+00&9.473E-03&3.295E-02\\
0.9679&2.083E+13&8.474E-04&1.094E-02&1.049E-02&1.539E-02&0.000E+00&0.000E+00&1.081E-02&3.767E-02\\
0.9706&1.867E+13&9.464E-04&1.263E-02&1.202E-02&1.729E-02&0.000E+00&0.000E+00&1.229E-02&4.289E-02\\
0.9732&1.673E+13&1.053E-03&1.519E-02&1.376E-02&1.933E-02&0.000E+00&0.000E+00&1.410E-02&4.933E-02\\
0.9755&1.499E+13&1.168E-03&1.821E-02&1.565E-02&2.148E-02&0.000E+00&0.000E+00&1.612E-02&5.651E-02\\
0.9776&1.344E+13&1.291E-03&2.145E-02&1.766E-02&2.368E-02&0.000E+00&0.000E+00&1.826E-02&6.408E-02\\
0.9795&1.204E+13&1.416E-03&2.568E-02&1.968E-02&2.584E-02&0.000E+00&0.000E+00&2.072E-02&7.262E-02\\
0.9812&1.079E+13&1.541E-03&3.015E-02&2.166E-02&2.791E-02&0.000E+00&0.000E+00&2.323E-02&8.126E-02\\
0.9828&9.674E+12&1.663E-03&3.416E-02&2.353E-02&2.985E-02&0.000E+00&0.000E+00&2.557E-02&8.921E-02\\
0.9842&8.665E+12&1.779E-03&3.766E-02&2.529E-02&3.165E-02&0.000E+00&0.000E+00&2.767E-02&9.637E-02\\
0.9854&7.757E+12&1.889E-03&4.018E-02&2.693E-02&3.331E-02&0.000E+00&0.000E+00&2.938E-02&1.023E-01\\
0.9866&6.943E+12&1.993E-03&4.193E-02&2.845E-02&3.484E-02&0.000E+00&0.000E+00&3.076E-02&1.072E-01\\
0.9876&6.214E+12&2.092E-03&4.315E-02&2.988E-02&3.625E-02&0.000E+00&0.000E+00&3.191E-02&1.114E-01\\
0.9886&5.566E+12&2.185E-03&4.406E-02&3.122E-02&3.757E-02&0.000E+00&0.000E+00&3.291E-02&1.150E-01\\
0.9895&4.991E+12&2.273E-03&4.498E-02&3.248E-02&3.881E-02&0.000E+00&0.000E+00&3.387E-02&1.185E-01\\
0.9903&4.482E+12&2.357E-03&4.627E-02&3.366E-02&3.996E-02&0.000E+00&0.000E+00&3.492E-02&1.222E-01\\
0.9911&4.033E+12&2.435E-03&4.798E-02&3.478E-02&4.103E-02&0.000E+00&0.000E+00&3.606E-02&1.262E-01\\
0.9918&3.639E+12&2.510E-03&4.943E-02&3.582E-02&4.203E-02&0.000E+00&0.000E+00&3.708E-02&1.298E-01\\
0.9925&3.290E+12&2.579E-03&5.046E-02&3.679E-02&4.296E-02&0.000E+00&0.000E+00&3.792E-02&1.328E-01\\
0.9931&2.981E+12&2.644E-03&5.045E-02&3.769E-02&4.381E-02&0.000E+00&0.000E+00&3.838E-02&1.346E-01\\
0.9937&2.708E+12&2.705E-03&4.942E-02&3.853E-02&4.460E-02&0.000E+00&0.000E+00&3.848E-02&1.353E-01\\
0.9942&2.465E+12&2.761E-03&4.791E-02&3.931E-02&4.534E-02&0.000E+00&0.000E+00&3.842E-02&1.353E-01\\
0.9947&2.250E+12&2.814E-03&4.631E-02&4.004E-02&4.602E-02&0.000E+00&0.000E+00&3.832E-02&1.352E-01\\
0.9952&2.059E+12&2.863E-03&4.491E-02&4.071E-02&4.664E-02&0.000E+00&0.000E+00&3.827E-02&1.351E-01\\
0.9957&1.889E+12&2.909E-03&4.377E-02&4.134E-02&4.723E-02&0.000E+00&0.000E+00&3.829E-02&1.352E-01\\
0.9961&1.736E+12&2.951E-03&4.320E-02&4.192E-02&4.776E-02&0.000E+00&0.000E+00&3.845E-02&1.358E-01\\
0.9964&1.599E+12&2.991E-03&4.347E-02&4.246E-02&4.826E-02&0.000E+00&0.000E+00&3.883E-02&1.372E-01\\
0.9968&1.474E+12&3.028E-03&4.475E-02&4.297E-02&4.872E-02&0.000E+00&0.000E+00&3.947E-02&1.395E-01\\
0.9971&1.360E+12&3.062E-03&4.750E-02&4.343E-02&4.914E-02&0.000E+00&0.000E+00&4.052E-02&1.431E-01\\
0.9974&1.256E+12&3.094E-03&5.212E-02&4.386E-02&4.953E-02&0.000E+00&0.000E+00&4.214E-02&1.486E-01\\
0.9977&1.160E+12&3.123E-03&5.917E-02&4.426E-02&4.989E-02&0.000E+00&0.000E+00&4.461E-02&1.564E-01\\
0.9980&1.070E+12&3.150E-03&6.888E-02&4.462E-02&5.022E-02&0.000E+00&0.000E+00&4.813E-02&1.669E-01\\
0.9982&9.852E+11&3.175E-03&8.197E-02&4.496E-02&5.052E-02&0.000E+00&0.000E+00&5.316E-02&1.806E-01\\
0.9984&9.046E+11&3.197E-03&9.911E-02&4.526E-02&5.079E-02&0.000E+00&0.000E+00&6.013E-02&1.984E-01\\
0.9986&8.266E+11&3.217E-03&1.204E-01&4.554E-02&5.104E-02&0.000E+00&0.000E+00&6.926E-02&2.202E-01\\
0.9988&7.502E+11&3.235E-03&1.452E-01&4.578E-02&5.125E-02&0.000E+00&0.000E+00&8.032E-02&2.454E-01\\
0.9989&6.751E+11&3.250E-03&1.720E-01&4.598E-02&5.143E-02&0.000E+00&0.000E+00&9.269E-02&2.727E-01\\
0.9990&6.019E+11&3.262E-03&2.003E-01&4.614E-02&5.158E-02&0.000E+00&0.000E+00&1.060E-01&3.013E-01\\
0.9991&5.340E+11&3.270E-03&2.295E-01&4.625E-02&5.167E-02&0.000E+00&0.000E+00&1.199E-01&3.307E-01\\
0.9992&4.739E+11&3.275E-03&1.744E-01&4.632E-02&5.173E-02&0.000E+00&0.000E+00&9.389E-02&2.758E-01\\
0.9992&4.239E+11&3.279E-03&4.945E-02&4.637E-02&5.178E-02&0.000E+00&0.000E+00&4.268E-02&1.509E-01\\
0.9992&3.881E+11&3.283E-03&4.003E-02&4.642E-02&5.182E-02&0.000E+00&0.000E+00&4.017E-02&1.416E-01\\
0.9993&3.524E+11&3.290E-03&3.167E-02&4.652E-02&5.191E-02&0.000E+00&0.000E+00&3.832E-02&1.334E-01\\
0.9995&3.067E+11&3.310E-03&2.935E-02&4.679E-02&5.215E-02&0.000E+00&0.000E+00&3.802E-02&1.316E-01\\
0.9997&2.780E+11&3.329E-03&4.248E-02&4.703E-02&5.237E-02&0.000E+00&0.000E+00&4.114E-02&1.452E-01\\
0.9999&2.713E+11&3.346E-03&2.949E-02&4.727E-02&5.257E-02&0.000E+00&0.000E+00&3.834E-02&1.327E-01\\
\end{tabular}
\label{tavolona}
\end{table}

\begin{figure}
\caption[a]{ Helioseismological determinations of the photospheric
Helium abundance, Y$_{ph}$. Same notation as in Table \ref{tab2}. The allowed
region according to the present work is between the dashed lines.}
\label{fig0}
\end{figure}

\begin{figure}
\caption[b]{The estimated global relative uncertainty on $U=P/\rho$,
according to the conservative approach, Eq. (\ref{cons}) (thick line), and to
the statistical approach, Eq. (\ref{stat}) (thin line).}
\label{fig1}
\end{figure}

\begin{figure}
\caption[dd]{ Helioseismological determinations and solar model predictions of 
properties of the outer sun. The box defines the region allowed by helioseismology.
Open circles denote models without diffusion, squares models with He diffusion, full circles
models with He and heavier elements diffusion.}
\label{fig2}
\end{figure}

\begin{figure}
\caption[eee]{ The difference between $U$ as predicted by selected solar 
models, $U_{mod}$ and the helioseismological determination, $U_{\odot}$,
 normalized to this latter. 
The allowed region is that within
the thick lines, corresponding to $\left ( \frac{\Delta U}{U}  \right ) _{con} $.
SUN24 is the ``model 0'' of Ref. 
\cite{ref10};
FRANEC96 is the ``best'' model 
with  He and heavier elements diffusion of
Ref. \cite{Ciacio};  BP95 is the model with metal and He diffusion of 
Ref. \cite{BP95}; JCD is the ``model S'' of Ref. \cite{sunjcd}.}
\label{fig3}
\end{figure}

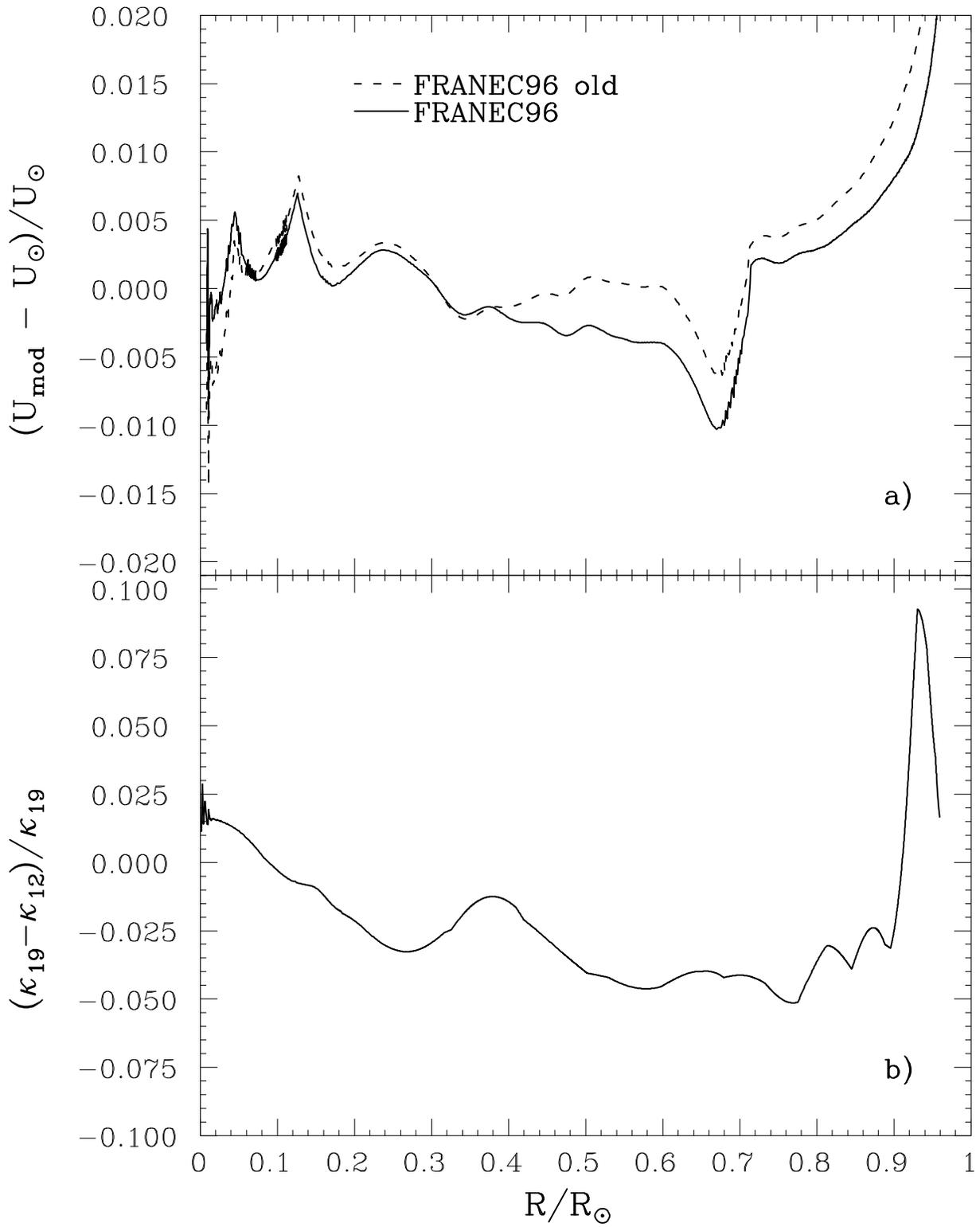
\begin{figure}
\caption[cc]{ a) The profiles of $U$ according to FRANEC96 and FRANEC96-OLD
(full and dashed lines, respectively).
b) The relative differences $\Delta k/k$ between 19- and 12- elements opacities, along
the solar profile.
}
\label{fig5}
\end{figure}

\end{document}